\documentclass[11pt]{article}

\usepackage[final]{acl}

\usepackage{times}
\usepackage{latexsym}
\usepackage{amsmath}

\usepackage[T1]{fontenc}

\usepackage[utf8]{inputenc}

\usepackage{microtype}

\usepackage{inconsolata}

\usepackage{graphicx}
\usepackage{listings}
\usepackage{booktabs}

\usepackage{tikz}
\usetikzlibrary{arrows.meta, positioning, fit, backgrounds, shapes.geometric, calc}

\definecolor{userblue}{RGB}{52, 101, 164}
\definecolor{localgreen}{RGB}{56, 142, 60}
\definecolor{remotered}{RGB}{183, 65, 53}
\definecolor{metricgold}{RGB}{160, 120, 10}
\definecolor{piicolor}{RGB}{183, 65, 53}
\definecolor{quasicolor}{RGB}{0, 117, 255}
\definecolor{okgreen}{RGB}{30, 130, 60}
\definecolor{warnred}{RGB}{183, 65, 53}
\definecolor{neutralgray}{RGB}{90, 90, 90}
\definecolor{arrowgray}{RGB}{120,120,120}

\tikzset{
  querybox/.style={
    rectangle, rounded corners=5pt,
    draw=#1!60!black, fill=#1!8,
    thick, inner sep=9pt,
    text width=5.6cm,
    align=left,
    font=\small\sffamily
  },
  headernode/.style={
    rectangle, rounded corners=3pt,
    fill=#1!75!black,
    inner sep=4pt,
    font=\footnotesize\bfseries\sffamily\color{white},
    text width=5.6cm,
    align=center
  },
  kanon/.style={
    rectangle, rounded corners=3pt,
    draw=#1!60!black, fill=#1!15,
    inner sep=4pt,
    font=\footnotesize\bfseries\sffamily,
    text=#1!50!black,
    minimum width=5.6cm,
    align=center
  },
  arrow/.style={-Stealth, thick, color=arrowgray},
}

\title{Beyond Direct Identifiers: Probabilistic Privacy Risk Estimation for\\Privacy-Conscious LLM Query Delegation}

\author{Li Siyan, Zhou Yu, Julia Hirschberg \\
  Department of Computer Science\\
  Columbia University\\
  \texttt{siyan.li@columbia.edu}  \\}

\usepackage[dvipsnames]{xcolor}

\begin{document}

\maketitle
\begin{figure*}[!ht]
    \centering
    \newcommand{\pii}[1]{{\color{piicolor}\underline{#1}}}
    \newcommand{\quasi}[1]{{\color{quasicolor}\textit{\underline{#1}}}}
    \resizebox{\textwidth}{!}{%
\begin{tikzpicture}[node distance=0.38cm]

\node[headernode=remotered] (h1)
  {(1) Raw Query};

\node[querybox=remotered, below=0.0cm of h1] (q1) {%
  ``I'm a \pii{34-year-old} \quasi{nurse} living in \pii{Austin, TX}.
  I was recently \quasi{diagnosed with Type 2 diabetes}
  and I'm \quasi{pregnant with my first child}.
  What dietary guidelines should I follow?''%
};

\node[kanon=warnred, below=0.0cm of q1] (k1)
  {$k$-anonymity $\approx$ \textbf{1} $\quad$ PII leaked: \textbf{2}};

\node[headernode=metricgold, right=0.5cm of h1] (h2)
  {(2) Naive PII Removal};

\node[querybox=metricgold, below=0.0cm of h2] (q2) {%
  ``I'm a \quasi{nurse} living in \textit{a city}.
  I was recently \quasi{diagnosed with Type 2 diabetes}
  and I'm \quasi{pregnant with my first child}.
  What dietary guidelines should I follow?''%
};

\node[kanon=metricgold, below=0.0cm of q2] (k2)
  {$k$-anonymity $\approx$ \textbf{980} $\quad$ PII leaked: \textbf{0}};

\node[headernode=localgreen, right=0.5cm of h2] (h3)
  {(3) PCD-$k$ Rewrite};

\node[querybox=localgreen, below=0.0cm of h3] (q3) {%
  ``Someone managing \textit{a chronic metabolic condition}
  during \textit{pregnancy} is looking for general
  dietary guidelines. What are the key
  nutritional recommendations?''%
};

\node[kanon=okgreen, below=0.0cm of q3] (k3)
  {$k$-anonymity $\approx$ \textbf{1.2M} $\quad$ PII leaked: \textbf{0}};


\node[below=0.3cm of k2, font=\small\sffamily, text=black,
      align=center] (legend) {%
  \textcolor{piicolor}{\underline{underlined red}} = PII \quad
  \textcolor{quasicolor}{\textit{\underline{underlined blue italic}}} = non-explicit quasi-identifier%
};

\end{tikzpicture}
}
    \caption{A motivating example demonstrating that eliminating explicit identifier leakage alone is not sufficient to fully preserve user privacy, as $k$-anonymity remains relatively low even after naive identifier removal. 
    }
    \label{fig:motivation}
\end{figure*}
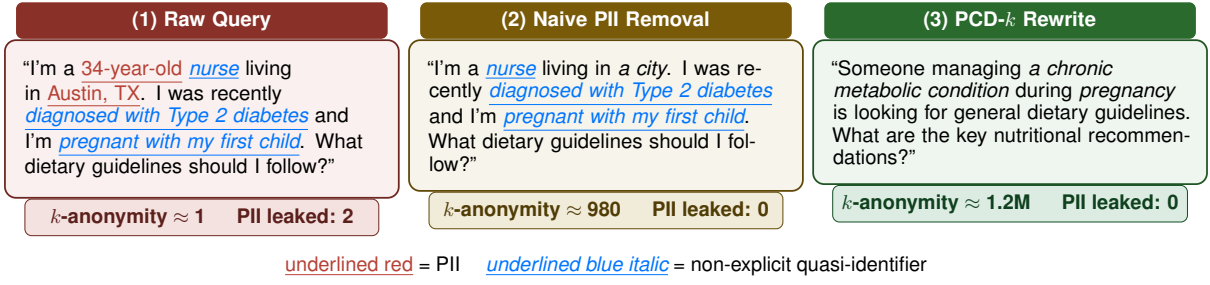

\begin{abstract}
    Recent work on protecting privacy during user-LLM interactions often focuses on direct, explicit identifiers: the personally-identifiable information (PII) captured by standard detectors. One such approach is \textbf{Privacy-Conscious Delegation (PCD)}, where a local LLM acts as an intermediary. However, privacy risk does not stem solely from explicit identifiers but also PII-free self-disclosures, leaving users identifiable through combinations of quasi-identifying traits. We investigate a probabilistic variant of PCD, where we augment its objectives with an LLM-driven probabilistic estimation of \textit{k-anonymity}. To facilitate this, we first create the \textbf{PUPA-SD} dataset, which contains naturalistic user queries with self-disclosure. Our preliminary results indicate that optimizing PAPILLON on PUPA-SD improves quality on unseen conversations across a variety of local models and produces the best privacy-utility balance for Llama-3.2-3B, while smaller models struggle to jointly optimize quality and privacy. We propose $k$-anonymity as a useful auxiliary metric for tackling PCD. 
\end{abstract}

\section{Introduction and Background}

LLM-powered agents and interactive systems pose significant privacy risks \cite{he2025emerged}, not only induced by \textbf{memorization} of user information \cite{carlini2022quantifying} \textbf{from model training} and therefore vulnerability to data extraction attacks \cite{carlini2021extracting}, but also by \textbf{self-disclosure during human-LLM interactions} \cite{mireshghallah2024trust,zhang2024privacyasst}. To exacerbate the matter, as LLMs exhibit more and more human-like traits in conversation, such as empathy and matching user linguistic styles \cite{peter2025benefits}, users develop greater trust in the models \cite{cohn2024believing,reinecke2025double}, increasing their willingness to share intimate personal details. Sharing private information increases \textbf{inference-time privacy risks}, since these self-disclosures could be used for the next round of training or divulged due to server-side data breaches. 

Prior work on protecting users from inference-time privacy risks has focused on reducing the exposure of PIIs and explicit identifiers to untrusted, remote frontier models through text sanitization and abstraction \cite{chowdhury2025pr,huang2024nap,pilan2025truthful} or by leveraging trusted, locally hosted models \cite{bae2025ppmi}. One framework that formalizes this task structure is \textbf{Privacy-Conscious Delegation (PCD)} \cite{siyan-etal-2025-papillon,hui2025privacypad}. PAPILLON serves as a strong baseline for PCD. We discuss PCD, PAPILLON, and the PUPA dataset from \citet{siyan-etal-2025-papillon} in Section \ref{sec:background}.

It is important to note, however, that eliminating explicit identifiers may not fully preserve anonymity: users can remain identifiable via \emph{quasi-identifiers} (traits whose combinations uniquely narrow the anonymity set). The notion of $k$-anonymity \cite{sweeney2002k} formalizes this risk in terms of the ease of pinpointing a person given a set of traits with respect to a known population. To operationalize anonymity reasoning for free-form texts, \citet{zhengprobabilistic} introduces BRANCH, a probabilistic framework that leverages LLMs' knowledge of census data and logical dependencies to estimate $k$-anonymity with relatively high accuracy. BRANCH therefore allows us to integrate $k$-anonymity into PAPILLON-based pipelines, leveraging the flexibility of prompt optimization. 

In this work, we take a first step toward \textbf{probabilistic privacy-conscious delegation} by augmenting PCD objectives beyond explicit identifier leakage.
Concretely, we begin with PAPILLON as a strong PCD framework and integrate BRANCH-based probabilistic \(k\)-anonymity as an additional objective. In order to optimize and evaluate our probabilistic framework, we extract \textbf{PUPA-SD} dataset from subsets of WildChat \cite{zhao2024wildchat} and LMSYS-Chat-1M \cite{zheng2023lmsys}, containing 166 real user queries containing self-disclosures. We show that optimizing for
$k$-anonymity on PUPA-SD can transfer to held-out PUPA-TNB, with \texttt{Llama-3.2-3B} achieving the strongest privacy-utility balance after optimization (Quality: 67.4, PII Leakage: 11.0).

\section{Background}
\label{sec:background}

\subsection{Privacy-Conscious Delegation}
\label{sec:pcd_detail}
In \textbf{PCD}, given a user query $q$ containing private information units $p_1, p_2, ..., p_n$, local model $M_{\text{\textsc{Local}}}$, and remote model $M_{\text{\textsc{Remote}}}$, a PCD system must ensure $M_{\text{\textsc{Remote}}}$ receives as little private information as possible while maintaining comparable response quality to passing $q$ into $M_{\text{\textsc{Remote}}}$ directly. 

\noindent \textbf{PAPILLON} provides a strong, prompt-optimization-based baseline for PCD. Given $q$, $M_{\text{\textsc{Local}}}$ produces a redacted query $q'$ that is sent to $M_{\text{\textsc{Remote}}}$, whose response is modified by $M_{\text{\textsc{Local}}}$ to cater to $q$. The pipeline can be prompt optimized, maintaining quality on 85.5\% of user queries while leaking only 7.5\% of information when using \texttt{Llama-3.1-8B-Instruct} as $M_{\text{\textsc{Local}}}$. PAPILLON presents one approach to addressing PCD, but it is not the only one; for instance, instead of firing a query to $M_{\text{\textsc{Remote}}}$ for \textit{every} user query, $M_{\text{\textsc{Local}}}$ could be more selective.

\noindent \textbf{PUPA} is created from extracting PII-containing user utterances from WildChat \cite{zhao2024wildchat}. We use its held-out set, PUPA-TNB, for evaluation. 

\subsection{BRANCH}
\label{sec:branch_detail}
Estimating $k$-anonymity over free-form text is difficult because disclosures are unstructured and statistically interdependent. BRANCH \citep{zhengprobabilistic} addresses this by modeling a text as a joint distribution over personal attributes. Given a text, it first selects the disclosures that can plausibly be estimated from population statistics (e.g., \textit{nurse}, \textit{Austin, TX}, \textit{pregnant}), then elicits
a Bayesian network over them: an LLM chooses an ordering of the disclosures as random variables and, for each one, determines which of the variables it is conditionally dependent on. This factors the joint distribution into probability terms, each of which is converted into a natural-language query and estimated by an LLM from its pretraining knowledge of demographic statistics. Recombining these estimates and multiplying by a population size yields the expected number of people matching the disclosed
attributes. Because the estimator consumes only text and returns a scalar, it can be directly incorporated into a prompt-optimization objective.

\section{PUPA-SD Dataset}
\label{sec:pupa_sd}

Following PUPA, the dataset used to construct and evaluate PAPILLON for PCD \cite{siyan-etal-2025-papillon}, we construct \textbf{PUPA-Self-Disclosure (PUPA-SD)} from initial conversation turns in WildChat and LMSYS-Chat-1M. For feasible manual inspection, we limit each dataset to its first 20,000 instances. 

\subsection{Self-Disclosure Extraction}
To quickly identify conversations with self-disclosure and improve the evaluation pipeline's accuracy in Section \ref{sec:pcd-k}, we develop a validated self-disclosure extraction module. We choose to create our own module rather than using existing disclosure extraction models \cite{dou2024reducing} because they are often trained on Reddit-style posts, which may represent a significant shift in distribution from conversation data.

Because the extraction process handles large data volumes, we need a fast model for quick evaluation and improvement. Therefore, we test three models with generally high performance and decent speed: \texttt{GPT-4.1-mini}, \texttt{GPT-5-mini}, and \texttt{GPT-4o-mini}. We define seed prompts for a self-disclosure extractor and evaluate each model on 50 synthetic texts with human-annotated disclosure from \citet{zhengprobabilistic} to identify the most suitable model. An \texttt{GPT-5}-based LLM judge is fixed to compare alternatives using the F-1 score to identify disclosures, concretely computed as follows:
\begin{align*}
    p &= \frac{\text{\# Correct Disclosures}}{\text{\# Predicted Disclosures}}\\
    r &= \frac{\text{\# Correct Disclosures}}{\text{\# Ground Truth Disclosures}}\\
    F_1 &= \frac{2 * p * r}{p + r}
\end{align*}

Results are documented in Table \ref{tab:disclosure_extrator}. Given its superior performance, we select \texttt{GPT-5-mini} for all disclosure extraction and ordering tasks (prompts in Appendix \ref{app:prompts}). We further optimize this module using DSPy's MIPROv2 \cite{opsahl2024optimizing} and \texttt{GPT-5} to generate 10 candidate prompt sets \cite{siyan2025passive}, selecting the best-performing combination \textbf{(Disclosure F-1 = 72.5)}.

\begin{table}[]
    \centering
    \begin{tabular}{c|ccc}
    \toprule
       \textbf{Model} & \textbf{Prec.} & \textbf{Rec.} & \textbf{F-1}  \\
    \midrule
       \texttt{GPT-4.1-mini} & \textbf{77.3} & 69.9 & 70.3 \\
       \texttt{GPT-5-mini} & 66.6 & \textbf{79.3} & \textbf{70.7} \\
       \texttt{GPT-4o-mini} & 74.6& 51.7 & 58.9 \\
    \bottomrule
    \end{tabular}
    \caption{Precision, recall, and F-1 scores of different models when evaluated on the synthetic data from \citet{zhengprobabilistic} using our LLM judge.}
    \label{tab:disclosure_extrator}
\end{table}


\subsection{Data Processing}
We apply the finalized extraction pipeline to the 40,000 initial conversation turns from WildChat and LMSYS-Chat-1M, and additionally filter out non-English and sexual instances during manual inspection. This process yielded a set of 166 turns. We aim to perform further processing and extraction in the future to grow the dataset size. See example instances of PUPA-SD in Appendix \ref{sec:pupa_sd_ex}.

\begin{figure}[!h]
    \centering
    \includegraphics[width=0.6\linewidth]{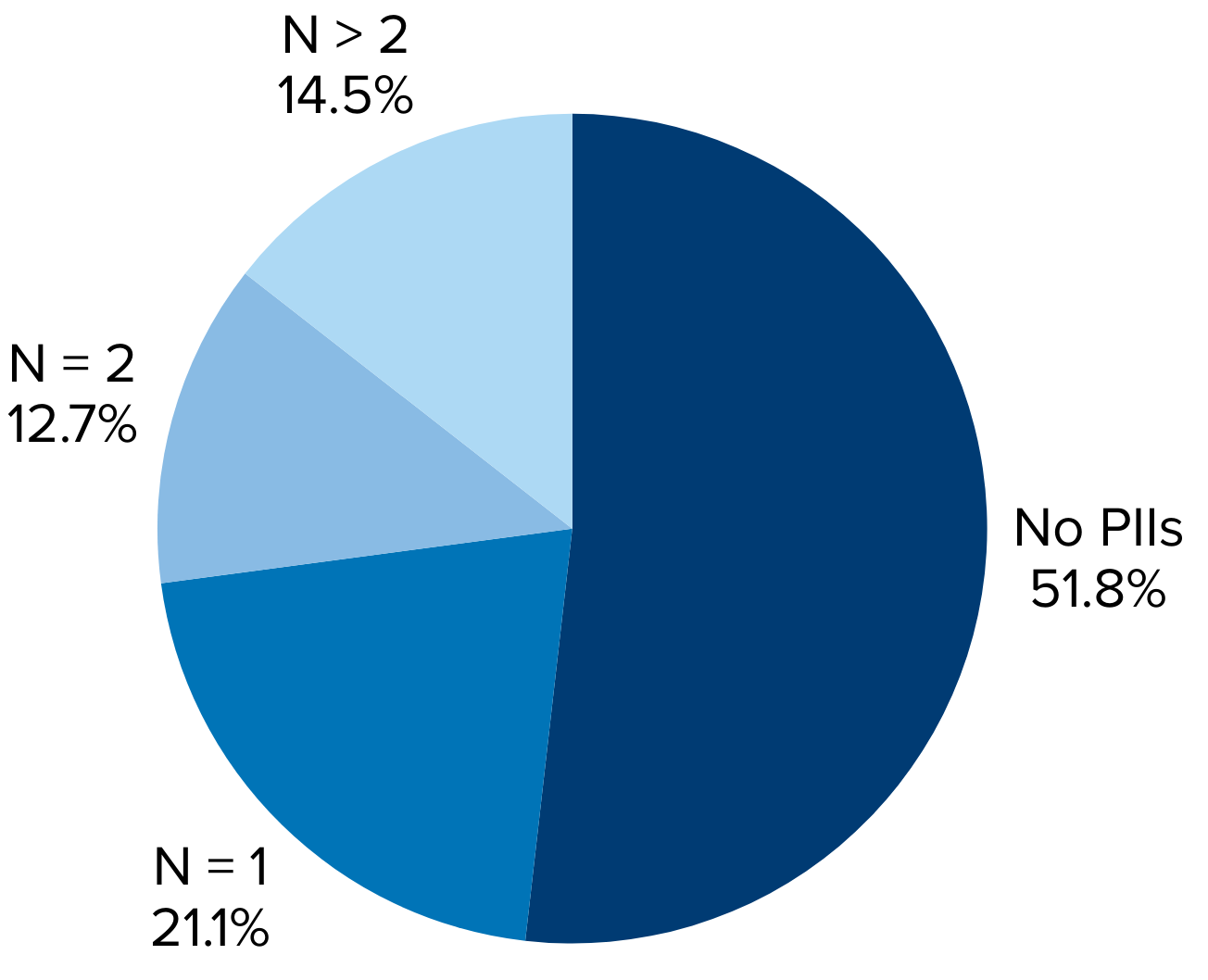}
    \caption{The distribution of the number of PII units in PUPA-SD instances. Here, N represents the number of PII units present, as defined by GLiNER-PII extractions.}
    \label{fig:pupa_sd_dist}
\end{figure}

PUPA-SD contains PII-free instances. To quantify this, we use GLiNER-PII\footnote{\url{https://huggingface.co/nvidia/gliner-PII}}, a state-of-the-art PII extraction model from NVIDIA, to extract information, including emails, names, company names, and locations, from PUPA-SD queries. We present the distribution of the number of extracted PIIs in Figure \ref{fig:pupa_sd_dist}. This further emphasizes that eliminating PII leakage does not fully prevent self-disclosure.

\section{Privacy-Conscious Delegation with $k$-Anonymity (PCD-$k$)}
\label{sec:pcd-k}
\subsection{Task Definition}
Similar to the original definition of Privacy-Conscious Delegation, our modified task involves utilizing both a trusted but weaker model, $M_{\textsc{Local}}$, and an untrusted but stronger one, $M_{\textsc{Remote}}$. $M_{\textsc{Local}}$ produces a query $q'$ based on the private user query $q$ containing PII units $p_1, p_2, ... p_n$. Here, in addition to preserving the quality of the final response with respect to the original response from the LLM, as well as reducing the amount of PII unit leakage to $M_{\textsc{Remote}}$, we additionally aim to increase $k$-anonymity of information passed to $M_{\textsc{Remote}}$. To evaluate our task, we employ the same validated LLM judges from \citet{siyan-etal-2025-papillon}; additionally, we implement a version of BRANCH to estimate $k$-anonymity.

\subsection{Implementing Probabilistic Risk Estimation}
BRANCH \cite{zhengprobabilistic} is a heavily LLM-dependent approach for estimating privacy risk. While accurate, it is costly, motivating improved data efficiency for swift \textit{evaluation} of our probabilistic PAPILLON pipelines.

We adopt BRANCH's disclosure ordering and query generation stages, simplifying the elicited Bayesian structure to a sequence of cumulative conditioning groups, in which disclosures entering the same group are treated as conditionally independent given the preceding group. We apply a minimally altered version (Appendix \ref{app:prompts}) of the self-disclosure extractor, modified to extract disclosures about \textit{any} individual. We then apply the LLM-based disclosure ordering and query generation from \citet{zhengprobabilistic}. \texttt{GPT-5-mini} estimates either the total originating population or the percentage of individuals with given attributes; if insufficient information exists, a fallback population of 400M (English speakers) is used. To reduce redundant LLM calls, queries and normalized answers are indexed using \texttt{openai/text-embedding-3-small}: a new query with cosine similarity $\geq 0.95$ to a stored query reuses its cached result. $k$-anonymity is then computed by sequentially multiplying the population estimate by each percentage, as in BRANCH.

 


\subsection{Evaluation Metric}
In addition to the \textbf{final response quality, PII leakage, and prompt well-formedness} metrics from \citet{siyan-etal-2025-papillon}, we instantiate a \textbf{$k$-anonymity metric}. To scale the $k$-anonymity metric to the [0,1] range in accordance with PAPILLON, we note that while dividing by the total population (400M) is a natural normalization, it yields near-zero values where meaningfully different anonymity levels (e.g., $k = 1,000$ vs. $k = 100{,}000$) become indistinguishable. We therefore define the reward as $\log_2(k) / \log_2(400\text{M})$, which compresses the range while preserving meaningful distinctions across anonymity levels.

\section{Experiments}

\subsection{$k$-Anonymity Pipeline Sanity Check}

To ensure our probabilistic k-anonymity estimator behaves as expected, we examine whether estimated k-anonymity \textit{negatively correlates with the degree of self-disclosure} in the original query. Intuitively, queries involving more PIIs should narrow the anonymity set and yield lower k-anonymity estimates. We compute the Spearman correlation between the number of PIIs per query and the estimated k-anonymity on PUPA-TNB, which contains a higher density of PII and self-disclosure than PUPA-SD, obtaining $\rho = -0.4045$ ($p < 10^{-12}$). We emphasize that this correlation is expected by construction: because the estimate is a function of the detected disclosures, a negative relationship confirms the pipeline's monotonic behavior but does \textit{not} establish that the estimations are calibrated against real statistics. We therefore treat this as a sanity check rather than formal validation, and discuss calibration as a limitation.


\subsection{Prompt Optimization}

We adopt PAPILLON as our PCD-$k$ pipeline unchanged, augmenting only the optimization objective to additionally account for k-anonymity. For prompt optimization, we apply DSPy's SIMBA\footnote{\url{https://dspy.ai/api/optimizers/SIMBA/}} prompt optimizer. Given a metric and a threshold of success, SIMBA samples minibatches of data, identifies difficult examples via metric variance, and then either updates the prompt via self-reflection or adds successful input-output pairs as in-context demonstrations. We limit the number of demonstrations to one.

Throughout our prompt optimization experiments, the metric is computed as follows:
$\text{Metric}_{\text{PAPILLON}} + k\_\text{Anon}(q')$, where $\text{Metric}_{\text{PAPILLON}}$ is a weighted sum of response quality and prompt well-formedness, i.e., the utility term. \textbf{PII leakage is not included during optimization, but we measure it during evaluation.} This composite metric value is normalized.

We optimize on PUPA-SD and evaluate on PUPA-TNB, testing {Llama-3.2-\{1B,3B\}-Instruct}, {Llama-3.1}{-8B-Instruct}, and {Qwen-2.5-}{\{0.5B,1.5B,7B\}-Instruct} as $M_{\textsc{Local}}$, with \texttt{GPT-4o-mini} fixed as $M_{\textsc{Remote}}$ and \texttt{GPT-5-mini} as the LLM judge.

\begin{table*}[!ht]
    \centering
    \begin{tabular}{lcccccc}
    \toprule
       & \multicolumn{3}{c}{\textbf{Before Optimization}} & \multicolumn{3}{c}{\textbf{After Optimization}} \\
    & \textbf{\textsc{Qual} $\uparrow$} & \textbf{\textsc{Leak} $\downarrow$} & \textbf{$k$-\textsc{Anon.} $\uparrow$} & \textbf{\textsc{Qual} $\uparrow$} & \textbf{\textsc{Leak} $\downarrow$} & \textbf{$k$-\textsc{Anon.} $\uparrow$} \\
    \midrule
    Llama-3.2-1B-Instruct  & 38.30 & 44.48 & 72.49 & 48.05 & 33.25 & 73.97 \\
    Llama-3.2-3B-Instruct  & 54.22 & 29.13 & 94.67 & 67.37 & \textbf{11.03} & \textbf{95.46} \\
    Llama-3.1-8B-Instruct  & 74.65 & \textbf{24.85} & \textbf{96.01} & \textbf{82.79} & 35.77 & 94.35 \\
    Qwen-2.5-0.5B-Instruct & 26.88 & 54.98 & 93.86 & 45.58 & 62.09 & 94.50 \\
    Qwen-2.5-1.5B-Instruct & 47.47 & 54.40 & 95.48 & 64.81 & 62.95 & 94.66 \\
    Qwen-2.5-7B-Instruct   & \textbf{83.41} & 35.12 & 93.68 & 79.72 & 66.51 & 89.38 \\
    \bottomrule
    \end{tabular}
    \caption{Average quality, PII leakage, and $k$-anonymity scores (0–100, higher is better) of PAPILLON pipelines using various local models and GPT-4o-mini as the proprietary model on PUPA-TNB, which is not used for prompt optimization.}
    \label{tab:results_1}
\end{table*}

\begin{figure}
    \centering
    \includegraphics[width=\linewidth]{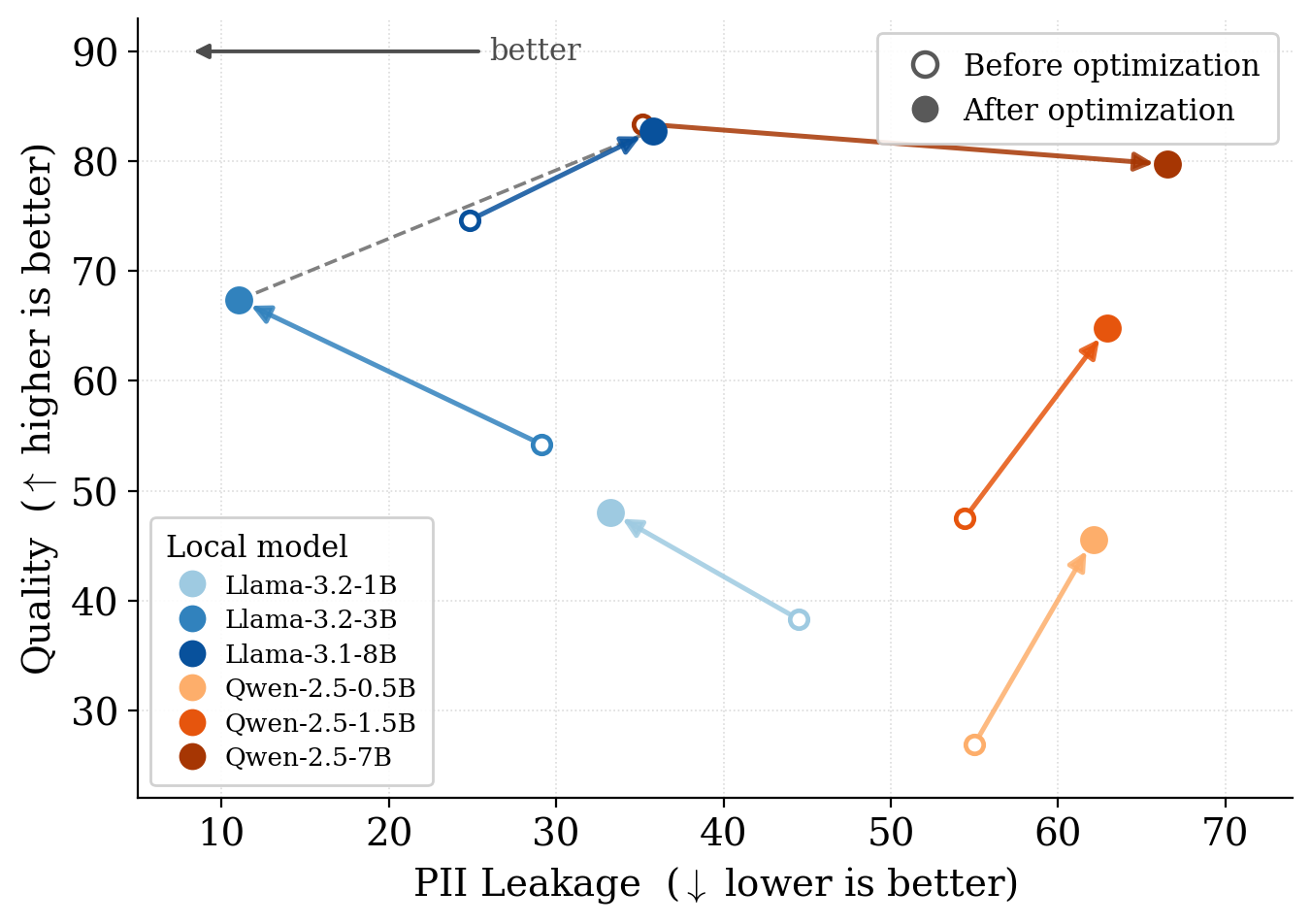}
    \caption{The Pareto line plot (before $\to$ after optimization on PUPA-SD) when evaluating models on PUPA-TNB. The dashed line establishes the post-optimization Pareto frontier. See the full results in Table \ref{tab:results_1}.}
    \label{fig:pareto_plot}
\end{figure}

Figure \ref{fig:pareto_plot} and Table \ref{tab:results_1} report results across all local models before and after SIMBA optimization on PUPA-SD, evaluated on the held-out PUPA-TNB. Optimization generally improves response quality, with the largest gains observed in \texttt{Llama-3.2-3B} (+13.2) and \texttt{Qwen-2.5-1.5B} (+17.3). PII leakage results are mixed: \texttt{Llama-3.2-3B} achieves a substantial reduction (29.1 $\to$ 11.0), while Qwen models show increased leakage after optimization, suggesting that smaller Qwen models struggle to jointly optimize quality and privacy. This trade-off is similarly observed for \texttt{Llama-3.1-8B-Instruct}, where the significant improvement in quality is accompanied by increased PII leakage. \texttt{Qwen-2.5-7B} is the only model where quality degrades after optimization (83.4 $\to$ 79.7), potentially due to the model already performing near its ceiling before optimization. $k$-anonymity scores remain largely stable across all models and conditions, which could be related to rewritten user queries either no longer containing disclosures or becoming impossible to estimate using publicly available data, resulting in the estimator falling back to the full population. One possible explanation of the heterogeneity of improvements across models is that, with the $k$-anonymity term contributing little variation and PII leakage excluded from the metric by construction, the objective is driven largely by the quality metric. We note that \texttt{Llama-3.2-1B} does receive a lower $k$-anonymity score (72.5), indicating that the metric remains discriminative when disclosures survive the rewrite, so this account is unlikely to apply uniformly. We leave a direct measurement of estimator fallback rates to future work.


\section{Discussion and Conclusion}

In this work, we take a first step toward probabilistic privacy-conscious delegation by augmenting PCD beyond PII leakage to incorporate $k$-anonymity as an objective. To support this, we introduce PUPA-SD and implement an efficient BRANCH-based k-anonymity estimator. Our experiments show that optimization on PUPA-SD yields improvements in quality for most models, but not all models successfully reduce leakage, suggesting that model capacity and benchmark headroom are key factors in navigating the privacy-utility tradeoff. Future work may examine the effect of optimizing on \textit{both} $k$-anonymity and PII leakage metrics. We hope this work motivates future research into probabilistic privacy risk estimation for human-LLM interaction, particularly as users increasingly share sensitive personal context with frontier models they do not control.



\section*{Limitation}

One core limitation is the potential inaccuracy in various parts of the evaluation pipeline. To begin with, BRANCH can often under-estimate $k$-anonymity as a result of the independence assumption. However, this is justifiable as the estimate would be relatively conservative, suitable for privacy preservation purposes. A more problematic issue is that our estimators are not calibrated against or grounded in real population statistics, as the estimation hinges entirely upon LLM-encoded prior knowledge. Due to financial constraints, we could not leverage the most capable models for self-disclosure extraction and k-anonymity estimation, meaning LLM-encoded census knowledge may be outdated or imprecise; future work could incorporate agentic web search to ground estimates in current data. 

PUPA-SD currently contains only 166 instances, limiting statistical power and precluding post-training approaches that could further boost smaller models' instruction-following ability; we plan to scale the dataset by broadening the extraction pipeline to additional conversations from both existing corpora and new sources. 

Finally, PUPA-SD and our evaluation pipeline are restricted to English; extending to multilingual settings would broaden applicability, though manual inspection would become increasingly challenging as more languages are incorporated.

\bibliography{custom}

\begin{thebibliography}{21}
\providecommand{\natexlab}[1]{#1}

\bibitem[{Bae et~al.(2025)Bae, Kim, Lee, Kim, Kim, Choi, and Mireshghallah}]{bae2025ppmi}
Yubeen Bae, Minchan Kim, Jaejin Lee, Sangbum Kim, Jaehyung Kim, Yejin Choi, and Niloofar Mireshghallah. 2025.
\newblock Ppmi: Privacy-preserving llm interaction with socratic chain-of-thought reasoning and homomorphically encrypted vector databases.
\newblock \emph{arXiv preprint arXiv:2506.17336}.

\bibitem[{Carlini et~al.(2022)Carlini, Ippolito, Jagielski, Lee, Tramer, and Zhang}]{carlini2022quantifying}
Nicholas Carlini, Daphne Ippolito, Matthew Jagielski, Katherine Lee, Florian Tramer, and Chiyuan Zhang. 2022.
\newblock Quantifying memorization across neural language models.
\newblock In \emph{The Eleventh International Conference on Learning Representations}.

\bibitem[{Carlini et~al.(2021)Carlini, Tramer, Wallace, Jagielski, Herbert-Voss, Lee, Roberts, Brown, Song, Erlingsson et~al.}]{carlini2021extracting}
Nicholas Carlini, Florian Tramer, Eric Wallace, Matthew Jagielski, Ariel Herbert-Voss, Katherine Lee, Adam Roberts, Tom Brown, Dawn Song, Ulfar Erlingsson, and 1 others. 2021.
\newblock Extracting training data from large language models.
\newblock In \emph{30th USENIX security symposium (USENIX Security 21)}, pages 2633--2650.

\bibitem[{Chowdhury et~al.(2025)Chowdhury, Glukhov, Anshumaan, Chalasani, Papernot, Jha, and Bellare}]{chowdhury2025pr}
Amrita~Roy Chowdhury, David Glukhov, Divyam Anshumaan, Prasad Chalasani, Nicolas Papernot, Somesh Jha, and Mihir Bellare. 2025.
\newblock Pr $\epsilon\epsilon$ mpt: Sanitizing sensitive prompts for llms.
\newblock \emph{arXiv preprint arXiv:2504.05147}.

\bibitem[{Cohn et~al.(2024)Cohn, Pushkarna, Olanubi, Moran, Padgett, Mengesha, and Heldreth}]{cohn2024believing}
Michelle Cohn, Mahima Pushkarna, Gbolahan~O Olanubi, Joseph~M Moran, Daniel Padgett, Zion Mengesha, and Courtney Heldreth. 2024.
\newblock Believing anthropomorphism: Examining the role of anthropomorphic cues on trust in large language models.
\newblock In \emph{Extended Abstracts of the CHI Conference on Human Factors in Computing Systems}, pages 1--15.

\bibitem[{Dou et~al.(2024)Dou, Krsek, Naous, Kabra, Das, Ritter, and Xu}]{dou2024reducing}
Yao Dou, Isadora Krsek, Tarek Naous, Anubha Kabra, Sauvik Das, Alan Ritter, and Wei Xu. 2024.
\newblock Reducing privacy risks in online self-disclosures with language models.
\newblock In \emph{Proceedings of the 62nd annual meeting of the association for computational linguistics (volume 1: long papers)}, pages 13732--13754.

\bibitem[{He et~al.(2025)He, Zhu, Ye, Liu, Zhou, and Yu}]{he2025emerged}
Feng He, Tianqing Zhu, Dayong Ye, Bo~Liu, Wanlei Zhou, and Philip~S Yu. 2025.
\newblock The emerged security and privacy of llm agent: A survey with case studies.
\newblock \emph{ACM Computing Surveys}, 58(6):1--36.

\bibitem[{Huang et~al.(2024)Huang, MacLean, Kang, Wu, Qu, Xu, Li, Yuan, and Haffari}]{huang2024nap}
Shuo Huang, William MacLean, Xiaoxi Kang, Anqi Wu, Lizhen Qu, Qiongkai Xu, Zhuang Li, Xingliang Yuan, and Gholamreza Haffari. 2024.
\newblock Napˆ 2: A benchmark for naturalness and privacy-preserving text rewriting by learning from human.
\newblock \emph{arXiv preprint arXiv:2406.03749}.

\bibitem[{Hui et~al.(2025)Hui, Dong, Sivapiromrat, Shareghi, and Collier}]{hui2025privacypad}
Zheng Hui, Yijiang~River Dong, Sanhanat Sivapiromrat, Ehsan Shareghi, and Nigel Collier. 2025.
\newblock Privacypad: A reinforcement learning framework for dynamic privacy-aware delegation.
\newblock \emph{arXiv preprint arXiv:2510.16054}.

\bibitem[{Mireshghallah et~al.(2024)Mireshghallah, Antoniak, More, Choi, and Farnadi}]{mireshghallah2024trust}
Niloofar Mireshghallah, Maria Antoniak, Yash More, Yejin Choi, and Golnoosh Farnadi. 2024.
\newblock Trust no bot: Discovering personal disclosures in human-llm conversations in the wild.
\newblock \emph{arXiv preprint arXiv:2407.11438}.

\bibitem[{Opsahl-Ong et~al.(2024)Opsahl-Ong, Ryan, Purtell, Broman, Potts, Zaharia, and Khattab}]{opsahl2024optimizing}
Krista Opsahl-Ong, Michael~J Ryan, Josh Purtell, David Broman, Christopher Potts, Matei Zaharia, and Omar Khattab. 2024.
\newblock Optimizing instructions and demonstrations for multi-stage language model programs.
\newblock In \emph{Proceedings of the 2024 Conference on Empirical Methods in Natural Language Processing}, pages 9340--9366.

\bibitem[{Peter et~al.(2025)Peter, Riemer, and West}]{peter2025benefits}
Sandra Peter, Kai Riemer, and Jevin~D West. 2025.
\newblock The benefits and dangers of anthropomorphic conversational agents.
\newblock \emph{Proceedings of the National Academy of Sciences}, 122(22):e2415898122.

\bibitem[{Pil{\'a}n et~al.(2025)Pil{\'a}n, Manzanares-Salor, S{\'a}nchez, and Lison}]{pilan2025truthful}
Ildik{\'o} Pil{\'a}n, Benet Manzanares-Salor, David S{\'a}nchez, and Pierre Lison. 2025.
\newblock Truthful text sanitization guided by inference attacks.
\newblock \emph{Applied Soft Computing}, page 114013.

\bibitem[{Reinecke et~al.(2025)Reinecke, Ting, Savulescu, and Singh}]{reinecke2025double}
Madeline~G Reinecke, Fransisca Ting, Julian Savulescu, and Ilina Singh. 2025.
\newblock The double-edged sword of anthropomorphism in llms.
\newblock In \emph{Proceedings}, volume 114, page~4. MDPI.

\bibitem[{Siyan et~al.(2025{\natexlab{a}})Siyan, Raghuram, Khattab, Hirschberg, and Yu}]{siyan-etal-2025-papillon}
Li~Siyan, Vethavikashini~Chithrra Raghuram, Omar Khattab, Julia Hirschberg, and Zhou Yu. 2025{\natexlab{a}}.
\newblock \href {https://doi.org/10.18653/v1/2025.naacl-long.173} {{PAPILLON}: Privacy preservation from {I}nternet-based and local language model ensembles}.
\newblock In \emph{Proceedings of the 2025 Conference of the Nations of the Americas Chapter of the Association for Computational Linguistics: Human Language Technologies (Volume 1: Long Papers)}, pages 3371--3390, Albuquerque, New Mexico. Association for Computational Linguistics.

\bibitem[{Siyan et~al.(2025{\natexlab{b}})Siyan, Zhang, Maharaj, Shi, and Li}]{siyan2025passive}
Li~Siyan, Jason Zhang, Akash Maharaj, Yuanming Shi, and Yunyao Li. 2025{\natexlab{b}}.
\newblock Is passive expertise-based personalization enough? a case study in ai-assisted test-taking.
\newblock \emph{arXiv preprint arXiv:2511.23376}.

\bibitem[{Sweeney(2002)}]{sweeney2002k}
Latanya Sweeney. 2002.
\newblock k-anonymity: A model for protecting privacy.
\newblock \emph{International journal of uncertainty, fuzziness and knowledge-based systems}, 10(05):557--570.

\bibitem[{Zhang et~al.(2024)Zhang, Xu, Ba, Wang, Hong, Liu, Qin, and Ren}]{zhang2024privacyasst}
Xinyu Zhang, Huiyu Xu, Zhongjie Ba, Zhibo Wang, Yuan Hong, Jian Liu, Zhan Qin, and Kui Ren. 2024.
\newblock Privacyasst: Safeguarding user privacy in tool-using large language model agents.
\newblock \emph{IEEE Transactions on Dependable and Secure Computing}, 21(6):5242--5258.

\bibitem[{Zhao et~al.(2024)Zhao, Ren, Hessel, Cardie, Choi, and Deng}]{zhao2024wildchat}
Wenting Zhao, Xiang Ren, Jack Hessel, Claire Cardie, Yejin Choi, and Yuntian Deng. 2024.
\newblock Wildchat: 1m chatgpt interaction logs in the wild.
\newblock \emph{arXiv preprint arXiv:2405.01470}.

\bibitem[{Zheng et~al.(2025)Zheng, Ritter, Das, and Xu}]{zhengprobabilistic}
Jonathan Zheng, Alan Ritter, Sauvik Das, and Wei Xu. 2025.
\newblock Probabilistic reasoning with llms for privacy risk estimation.
\newblock In \emph{The Thirty-ninth Annual Conference on Neural Information Processing Systems}.

\bibitem[{Zheng et~al.(2023)Zheng, Chiang, Sheng, Li, Zhuang, Wu, Zhuang, Li, Lin, Xing et~al.}]{zheng2023lmsys}
Lianmin Zheng, Wei-Lin Chiang, Ying Sheng, Tianle Li, Siyuan Zhuang, Zhanghao Wu, Yonghao Zhuang, Zhuohan Li, Zi~Lin, Eric~P Xing, and 1 others. 2023.
\newblock Lmsys-chat-1m: A large-scale real-world llm conversation dataset.
\newblock \emph{arXiv preprint arXiv:2309.11998}.

\end{thebibliography}

\appendix

\section{Prompts}
\label{app:prompts}
All prompts used for this work are written as DSPy modules. We use the latest version of DSPy (v3.1.2).
\subsection{Self-Disclosure Extraction}

Two versions of the self-disclosure extractor are used: one for extracting data from WildChat and LMSYS-Chat-1M, and the other for extracting self-disclosures from prompts created during the PAPILLON optimization and evaluation process. 

The main difference between the two versions is the emphasis on whether the disclosures are \textbf{self}-disclosures or disclosures about any person. The reason we make this design decision is as follows: When going through conversations in the datasets, it is critical that the extracted disclosures are about the user themselves, as these conversations often include fictional scenarios and characters that could be involved in sexual content. However, when it comes to extracting disclosures from prompts created by PAPILLON pipelines, as we are certain that the original user queries contain \textit{self-disclosures about the user}, we can be more liberal in our definition of disclosures and include all disclosures present in the text. We present the second version of the extractor below.

\lstset{
    basicstyle=\ttfamily\small,   
    breaklines=true,             
    breakatwhitespace=true,
    columns=fullflexible,
    keepspaces=true,             
    showstringspaces=false,
}
\begin{lstlisting}[language=Python]
class AnyDisclosureExtractorSignature(dspy.Signature):
    """From the text below, extract only those personal disclosures about ANY person that can be estimated using official, population-level statistics (e.g., census, national surveys, government records). Return a compact, canonical list of disclosures with one best-fit category per item.

Key definitions
- Personal Disclosure: Information about ANY person revealed in the text - including the author/poster, a described subject the request is written on behalf of (e.g., "a nurse who lives in...", "a third-year student"), or any named or role-described individual. Includes static traits (e.g., age, gender, location) and time-bound events (e.g., health diagnosis, education credential, employment change). The person does NOT need to be the one writing the text.
- Estimable: Attributes/events that are commonly measured in official data sources (census, government surveys/registries, administrative records). If it's not plausibly supported by reputable population statistics, exclude it.

Allowed categories (pick the single best fit for each item)
- location, age, relationship_status, gender, pet, appearance, race/nationality,
  sexual_orientation, health, family, occupation, mental_health, emotions,
  reproductive_health, finance, education, crime, events, PII, other

Inclusions (examples are illustrative, not exhaustive)
- Demographics: age ("25 year old"), gender ("female", "trans woman"), race/nationality ("Black", "Mexican American").
- Geography: residence or stable presence ("New York City", "California", "UK"). Prefer official place names; expand common unambiguous abbreviations (e.g., "NYC" $\to$ "New York City").
- Household/family: "married", "single", "divorced", "have 2 children", "live alone".
- Education: "high school graduate", "bachelor's degree", "college student", "in graduate school".
- Occupation and employment status: job titles ("software engineer"), statuses ("unemployed", "retired"), industry ("construction worker"). Freelance/gig counts as occupation; layoffs count only if stated as an event ("laid off").
- Health (physical): diagnoses, chronic conditions, disability status, medication use, BMI/height/weight if stated, substance use if specific.
- Mental health: diagnosed or clinically recognizable conditions ("diagnosed with depression", "ADHD"). Suicidal ideation/attempts belong here if explicitly stated.
- Reproductive health: pregnancy, contraception use/duration, fertility treatment, miscarriages, abortions.
- Finance: income bracket, rent amount, debt/loans, benefits/assistance participation.
- Pets: "pet dog", "2 cats".
- Crime/justice: "arrested for DUI", "felony conviction".
- Events (only if plausibly measured in official stats and not better placed above): "moved states this year", "evicted", "naturalized this year".
- PII (only if commonly available in official aggregate data): first name, ZIP code, year of birth. Exclude unique identifiers (full address, phone numbers, SSN, email).

Exclusions
- Purely incidental third parties with no representational role (e.g., celebrities or historical figures mentioned in passing, a researcher being contacted).
- Vague emotions/opinions ("I feel sad", "I'm overwhelmed") unless clearly tied to a diagnosable/estimable construct. If present without diagnosis, omit.
- Non-English disclosures: if the disclosure is not in English, omit it.
- Duplicates or near-duplicates ("woman" vs "female"; keep one).
- Unverifiable speculation ("might be pregnant", "maybe have ADHD") -- instead capture concrete, stated facts (e.g., "missed period" as health).

Normalization rules
- Keep spans short and canonical: "I'm 25" --> "25 year old"; "live in NYC" --> "New York City".
- Parse compact tokens: "18F" --> two items: "18" (age), "female" (gender).
- Prefer specific standard geographies (city + state/country) when given; do not invent detail.
- Map "student" to education; job titles and employment status to occupation; income/benefits to finance; relationship labels ("engaged", "married", "single", "widowed") to relationship_status.
- If multiple categories could apply, choose the best single fit. Do not duplicate across categories.
- Include durations when they are commonly measured (e.g., "on birth control for 3 years" under reproductive_health or health depending on context).

Edge cases
- If no eligible disclosures exist, return an empty list: <list></list>
- If both current and past facts are present, include both if clearly stated and estimable (e.g., "graduated last year" + "currently unemployed").
- When a subject is described in the third person but the request is clearly written on their behalf (e.g., "write a bio for a 35-year-old nurse in Boston"), extract their disclosures as you would for a first-person statement.

Output format (strict)
- Return ONLY the disclosures in this XML-like structure:
  <list>
    <answer>disclosure phrase</answer><type>category</type>
    ...
  </list>
- Do not include any other text. One disclosure per <answer>...</answer><type>...</type> pair.

Mini-examples
- Text: "I'm a 30M living in LA, recently laid off, now looking for work."
  Answer: <list><answer>30</answer><type>age</type>
  <answer>male</answer><type>gender</type>
  <answer>Los Angeles</answer><type>location</type>
  <answer>laid off</answer><type>events</type>
  <answer>unemployed</answer><type>occupation</type>
  </list>
- Text: "Write a professional email from a third-year clinical psychology student focusing on neuropsychology."
  Answer: <list><answer>third-year clinical psychology student</answer><type>education</type><answer>focusing
  on neuropsychology</answer>
  <type>education</type></list>
- Text: "I want to understand the cost of living in Northampton for a nurse who lives in a small studio apartment."
  Answer: <list><answer>nurse</answer><type>occupation</type>
  <answer>Northampton</answer><type>location</type>
  </list>

Now analyze the text and return the list. Only include disclosures that are estimable, normalized, non-duplicative, English, and mapped to the best-fit category."""
    text: str = dspy.InputField(description="The text to analyze for personal disclosures.")
    answer: str = dspy.OutputField(description="The list of personal disclosures in the text, formatted as specified above. Each disclosure should be short and succinct.")

\end{lstlisting}

\subsection{Disclosure Ordering}
\begin{lstlisting}[language=Python]
class DisclosureOrderingSignature(dspy.Signature):
    """**Task**: Arrange disclosures into cumulative conditioning groups so that the joint \
probability P(d1, d2, ..., dn) can be factored as a product of conditional probabilities \
(chain rule / BRANCH framework).

**Group construction algorithm**:
1. Identify INDEPENDENT disclosures -- those with no statistical dependencies \
(typically location, age, gender). These seed Group 1.
2. Identify FIRST-WAVE disclosures -- those that depend ONLY on the independent disclosures \
from step 1 (e.g. age given location+gender). Add them to Group 1 as well.
   --> Group 1 = independent disclosures + first-wave disclosures.
3. For each subsequent group: find all remaining disclosures whose EVERY dependency is \
already present in the previous group. Add them to form the next group.
   --> Group k = ALL items from Group k-1 PLUS the newly resolved disclosures.
4. Repeat until all disclosures are placed.

**Critical format rule**: every group is CUMULATIVE -- it repeats every disclosure from \
all prior groups and then appends the new ones. Do NOT create a new group for items that \
can be resolved within the current group's context.

**Example**:
Disclosures: New York City (location), woman (gender), 25 year old (age), software engineer (occupation)
- Independent: New York City, woman --> seed Group 1
- First-wave: 25 year old (depends only on location + gender) --> also in Group 1
- Group 1 = {New York City, woman, 25 year old}
- Resolved by Group 1: software engineer (depends on location, gender, age -- all in Group 1)
- Group 2 = Group 1 + {software engineer} = {New York City, woman, 25 year old, software engineer}

Output:
<list>\
<group><answer>New York City</answer><type>location</type>
<answer>woman</answer><type>gender</type>
<answer>25 year old</answer><type>age</type>
</group>\
<group><answer>New York City</answer><type>location</type>
<answer>woman</answer><type>gender</type>
<answer>25 year old</answer><type>age</type>
<answer>software engineer</answer><type>occupation</type>
</group>\
</list>

**Instructions**:
Return the result as <list></list> of <group> elements. \
Each group contains ALL disclosures accumulated up to that step: \
<group><answer>span1</answer><type>cat1</type>
<answer>span2</answer><type>cat2</type>...</group>
Do NOT omit any disclosure from any group. Do NOT create single-item groups unless only one \
disclosure exists at that step."""
    disclosures: str = dspy.InputField(description="The list of personal disclosures to reorder.")
    answer: str = dspy.OutputField(description="The reordered list of personal disclosures, formatted as specified above.")

\end{lstlisting}

\section{PUPA-SD Examples}
\label{sec:pupa_sd_ex}
We underline the self disclosures and write the corresponding disclosure category in bracketed red texts for the following examples.
\subsection{Example 1}
\begin{itemize}
    \item \textbf{User Query:} Write me a bio \\My Name is \underline{NAME\_1} {\color{red} [PII]} I am a \underline{student} {\color{red}[education]} who is currently a \underline{begginer free lancer} {\color{red} [occupation]} i provide services as\\ \underline{logo designing} {\color{red} [occupation]}\\ \underline{Social Media Manager} {\color{red} [occupation]}
    \item \textbf{Target Response:} NAME\_1 is a talented student with a passion for visual design and marketing. As a beginner freelancer, he offers a range of services including logo design and social media management. With his creative eye and strong communication skills, NAME\_1 is able to bring a unique perspective to each project he takes on. He is always eager to learn and grow as a professional, and is committed to delivering high-quality work to his clients. In his free time, NAME\_1 enjoys exploring new creative pursuits and staying up-to-date on the latest trends in design and marketing.
\end{itemize}

\subsection{Example 2}
\begin{itemize}
    \item \textbf{User Query:} I \underline{have a} website for a \underline{piano store} {\color{red} [occupation]}. I need to write SEO optimized content that is very engaging, and has local keywords in it for the \underline{Nashville} {\color{red} [location]} and the \underline{Middle Tennessee area} {\color{red} [location]}.
    \item \textbf{Target Response:} Sure, I can help you with that! To create SEO optimized content that is engaging and localized for Nashville and Middle Tennessee, here are some tips to keep in mind:
    
    1.   Use local keywords: Incorporate local keywords related to Nashville and Middle Tennessee into your content. This will help your website show up in local search results and attract people looking for piano stores in the area.\\
    2.   Use long-tail keywords: In addition to local keywords, use long-tail keywords that are more specific to your business. For example, if you're a piano store in Nashville, use keywords like ""best piano store in Nashville"" or ""piano lessons in Nashville.""\\
    3.   Use quality content: High-quality content is essential for SEO. Make sure your content includes valuable information about your piano store and the products and services you offer. This will help establish your website as an authority in the piano industry and attract more visitors.\\
    4.   Use meta descriptions: Meta descriptions are brief snippets of text that appear under each of your website's pages in search results. Include your local keywords and information about the content of the page in your meta descriptions.\\
    5.   Use internal linking: Internal linking is the practice of linking to other pages on your website from within your content. This can help search engines understand the structure of your website and improve your website's SEO.\\
    6.   Use alt tags: Alt tags are short descriptions of the images on your website. Include your local keywords and information about the content of the image in your alt tags.\\
    7.   Use social media: Use social media to promote your website and reach a wider audience. Share your content on your social media platforms and encourage your followers to share your content with their own followers.
    
    I hope these tips help you create engaging and localized content for your piano store website! Let me know if you have any other questions or if there's anything else I can help with.
\end{itemize}

\subsection{Example 3}

\begin{itemize}
    \item \textbf{User Query:} What is a good “doing business as” name for a \underline{13 year old} {\color{red} [age]} \underline{girl} {\color{red} gender} who \underline{makes TikTok videos about k-pop dances} {\color{red} [occupation]}?
    \item \textbf{Target Response:} Here are a few suggestions for a ""doing business as"" name for a 13-year-old girl who makes TikTok videos about K-Pop dances:\\
    1. KPop Dance Sensation\\
    2. K-Pop Kween\\
    3. K-Pop Kick\\
    4. K-Pop Fusion\\
    5. K-Pop Groove\\
    6. K-Pop Beats\\
    7. K-Pop Frenzy\\
    8. K-Pop Dancefloor\\
    9. K-Pop Fever\\
    10. K-Pop Starstruck
    
    It's important to choose a name that reflects your brand and resonates with your target audience. Consider what sets you apart from other K-Pop dance content creators and use that as inspiration for your brand name.
\end{itemize}

\end{document}